# GA-MSSR: Genetic Algorithm Maximizing Sharpe and Sterling Ratio Method for RoboTrading

Zezheng Zhang
School of Computer Science
The University of Sydney
Sydney, Australia
zaczhang1994@gmail.com

Matloob Khushi
School of Computer Science
The University of Sydney
Sydney, Australia
mkhushi@uni.sydney.edu.au

*Abstract*— Foreign exchange is the largest financial market in the world, and it is also one of the most volatile markets. Technical analysis plays an important role in the forex market and trading algorithms are designed utilizing machine learning techniques. Most literature used historical price information and technical indicators for training. However, the noisy nature of the market affects the consistency and profitability of the algorithms. To address this problem, we designed trading rule features that are derived from technical indicators and trading rules. The parameters of technical indicators are optimized to maximize trading performance. We also proposed a novel cost function that computes the risk-adjusted return, Sharpe and Sterling Ratio (SSR), in an effort to reduce the variance and the magnitude of drawdowns. An automatic robotic trading (RoboTrading) strategy is designed with the proposed Genetic Algorithm Maximizing Sharpe and Sterling Ratio model (GA-MSSR) model. The experiment was conducted on intraday data of 6 major currency pairs from 2018 to 2019. The results consistently showed significant positive returns and the performance of the trading system is superior using the optimized rule-based features. The highest return obtained was 320% annually using 5-minute AUDUSD currency pair. Besides, the proposed model achieves the best performance on risk factors, including maximum drawdowns and variance in return, comparing to benchmark models. The code can be accessed at https://github.com/zzzac/rule-based-forex-trading-system

## I. INTRODUCTION

Foreign exchange, abbreviated as forex, is the largest financial market in the world with over $5 trillion dollars transactions each day and it runs continuously from Monday to Friday, 24 hours a day. Forex is the most liquid financial market where little transaction fees are applied depending on brokers. The role of technical analysis had long been regarded as more than important in the field of foreign exchange [1]. It is widely believed that the pattern of price movement would repeat itself. Many algorithms are based on technical indicators with carefully selected parameters according to professional experience.

Many papers focused on the time series prediction problem. Under the growing influence of artificial intelligence, deep learning techniques were used to predict future price movements with carefully designed neural networks. Long Short-Term Memory (LSTM) is the state-of-art sequence learning network to forecast the stock price [2], and it could also be used to identify some common patterns in traded stocks as proposed by Fischer et al. [3]. Multi-task recurrent neural networks were used for stock price forecasting [4] which had better performance than LSTM on selected datasets. Selvin et al. [5] compared Convolutional Neural Networks (CNN) and LSTM architectures for predicting the stock price. Zeng and Khushi combined wavelet denoising and Attention-based RNN-ARIMA Model to predict USDJPY prices [6]. Bao et al. [7] used stacked autoencoder to extract high-level features of the price pattern for next day close price prediction. Gao [8] used LSTM and hourly stock prices to predict the trend. Support Vector Regressor is another common approach used for stock price prediction [9]. A combination of statistical models and deep learning models was proved to perform well on short term trend forecasting [10]. Zhang et al. [11] implemented a novel network, State Frequency Memory (SFM) to capture the multi-frequency trading patterns by utilizing Fourier Transform. However, simply predicting the next day price is not enough to consistently make profits unless the prediction is accurate enough when predicting multiple timesteps. To the best of our knowledge, a model with such adequate predictive power in the foreign exchange market has not been found.

CNN was also commonly used for classification tasks to identify the historical patterns on the chart and classify the trend. Tsai [12] classified the market into uptrend, sideways and downtrend and used CNN to classify the trend direction. The accuracy they achieved was not ideal possibly because of the lack of training data. Kusuma et al. [13] claimed they achieved 90% bi-directional accuracy using CNN with candlestick charts on a small dataset. Sezer et al. [14] combined 15 technical indicators in 15 timesteps into 2-D images and used CNN for classification. The model outperformed LSTM, MLP over a long period. The classification problem approach presents more value comparing to time series prediction problem since it would give signals when the trend changes. Nevertheless, the high volatility nature of the forex market sets a high barrier for machine learning models to outperform professional traders. A complete trading system needs to consider both prediction and risk management to maximize performance.

There are a few attempts to design an intelligent trading system that focus on strategies that optimize the overall trading performance. The applicability and efficiency of genetic algorithm in portfolio optimization was illustrated by Sefiane, et al. [15]. Mendes et al. [16] proposed a trading system using genetic algorithm and technical indicators to maximize the Stirling ratio but the performance on testing data struggle to make profits. Ozturk et al. [17] used heuristic based trading rules together with genetic algorithm which showed genetic algorithm could be used to select the best trading rules. Evans et al. [18] introduced decision making model using artificial neural networks and genetic algorithms and achieved 23.3% annualized net return. Maknickienė et al. [19] used LSTM-based neural networks with genetic algorithm as learning algorithm which resulted in significant increase of the reliability of prediction. Pawel et al. [20] developed decision trees using evolutionary algorithms and technical indicators to identify buy or sell signals. Rodrigo et al. [21] showed that genetic algorithm could also be used to optimize the parameters in technical indicators used for SVR+GHSOM model based trading system and the results outperformed the market. Bernardo et al. [22] categorized the market into three different types and used a hybrid system with SVM and genetic algorithm to classify the trend. The claimed annualized return is 83% with high leverage. Genetic algorithm is an example of an evolutionary computation method, which is capable of finding the near-global optimal solution of a non-linear non-convex function without getting trapped in local minima[23]. Petropoulos et al. [24] proposed using correlations between currency pairs as additional trading signals which were aggregated using the genetic algorithm and constrained optimization methods. The results showed claimed 17% annualized return with low leverage.

Technical indicators could reveal underlying information regarding the historical price movements. Hence, it is widely used for algorithm trading. However, given there are thousands of indicators do not always work since it is publicly available to everyone. The study attempts to design feature engineering process using trading rules, which are combinations of technical indicators. The features utilize crossover rules to models the relationship between technical indicators. The parameters for the technical indicators are selected using optimization methods with a focus on trading performance.

It is noted that many of the research papers only included annualized returns in their results. However, in any financial market, risk management is the crucial factor to successful and consistent returns. According to Capital Market Line (CML) theory, the excess return and the standard deviation of the return should be proportional as in Fig. 1 and the slope of the curve is also called the Sharpe ratio. Ideally, the best strategy is the one with the highest Sharpe ratio because it will have the highest expected return at a given risk level. Given the high volatility of forex market, the unexpected drawdown should also be considered. Some papers focused on the risk-adjusted returns using the Sharpe ratio or the Sterling ratio. Sharpe ratio penalizes negative returns less when the average return is generally low, and it would also penalize occasional high returns. Sterling ratio penalizes the maximum drawdown which could be experienced when market changes suddenly and unexpectedly, which is very common in forex market.

Due to the above deficiency of the performance measure, we propose a novel ratio which is essentially a combination of modified Sharpe Ratio and Stirling Ratio (SSR). The proposed ratio measures the risk-adjusted return that attempt to address the deficiency of the original ratios. We also propose a RoboTrading system, the Genetic Algorithm Maximizing SSR (GA-MSSR) model to build a trading strategy with trading rule features.

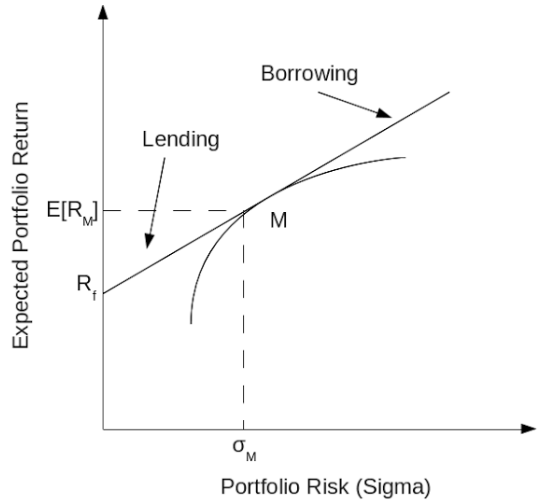

Fig. 1. Overview of the workflow

## II. PROPOSED METHODOLOGY

### A. Overview of process

The overall workflow of the forex trading system proposed in this paper is demonstrated in Fig. 2. The trading system starts with historical data feed and generates technical indicators using historical open, high, low and close prices. Feature engineering using trading rules are computed and the parameters for the rule-based features optimized using a cost function that computes the risk-adjusted return. The new features are fed to the genetic algorithm module to compute the weighting of the features that maximize the overall risk-adjusted return using the designed cost function.

### B. Technical indicators

Technical indicators are defined to represent the higher-level interpretation of past price movements. Each technical indicator is computed from the open, high, low and close prices with user-defined parameters. Three types of technical indicators are introduced.

1) Trend indicators
    a) Moving average
    b) Exponential moving average
    c) Double exponential moving average
    d) Triple exponential moving average
    e) Vortex indicators
2) Momentum indicators
    a) Relative strength index
    b) Stochastic oscillators
3) Volatility indicators
    a) Bollinger bands
    b) Ichimoku indicators

*c) Keltner channel*

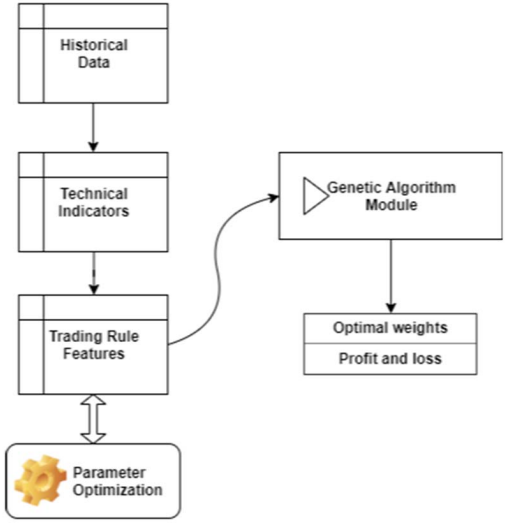

Fig. 2. Overview of the workflow

Volume is another important category of technical indicators. However, due to the decentralization nature of forex market, we cannot obtain accurate volume data. Therefore, volume indicators are not used.

Each technical indicator is essentially a time series with no future information. The technical indicators are essential for building the trading rules. It is inferred that trend, momentum and volatility information regarding to the past price movements could be used to predict the future price movement directions and the embedding risks.

*C. Trading rule derived feature engineering*

Crossover is one of the most important and common trading rules that are used by traders. Crossover rules refers to price sensitive movements when there is a crossing between technical indicators, open or close prices or a threshold of a constant value. Such events could indicate a change in the trend, momentum or volatility that would reveal important information for trading performance optimisation. Logical conditional operations will be used to generate features therefore the created feature only has 3 values, where 1 represents long indication, 0 represents a neutral indication and -1 indicates a short indication. The benefit of using simple representations as features is to remove noisy short-term fluctuations and yield higher-level interpretation that could potentially improve the performance of the machine learning algorithms for the trading system. The possible combination set of technical indicators is infinity. Hence, only 16 trading rules based on technical indicators were formulated and each of the 16 trading rules falls into one of the following 4 categories.

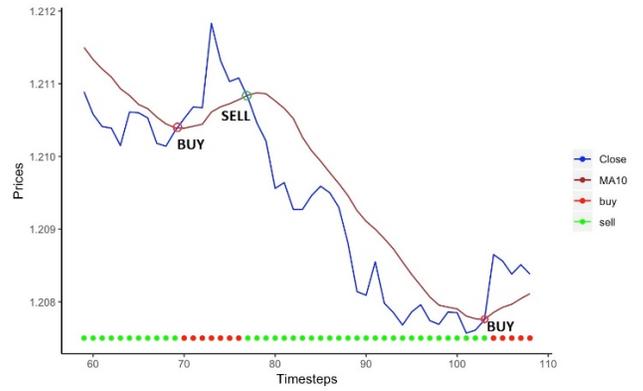

Fig. 3. Example of simple moving averages crossover rule

*1) Two time series crossing*

When two time series cross each other, the trading signal is inverted depending on the direction of the intersection. For example, if a slow moving average crosses above a fast moving average, the signal will be inverted from 1 to -1. Boolean operation is used to compare the values of the moving averages of the close price. In the two series crossing category, no neutral position or 0 value will occur.

$$signal_t = \begin{cases} 1 \text{ if } series1_t > series2_t \\ 0 \text{ else} \end{cases} \quad (1)$$

An example of two time series crossing rule is the simple moving average crossing rule. Two simple moving average time series need to be created where typically one moving average has a longer period and another one has a shorter period. A simple moving average on close price of period 10 (MA10) and a simple moving average of 1, which is the close price itself are shown in Fig. 3. When MA10 crosses close price from above, a buy signal is generated, and a sell signal is generated when close price cross MA10 from above. Between those intersections, the previous signal would persist representing no change in the position when there is no crossover. Therefore, using this strategy results in 100% of the capital invested all the time throughout the simulation.

Besides the simple moving average crossover rule, exponential moving average, stochastic oscillator and other technical indicators could be useful under category 1.

*2) One time series and a threshold crossing*

Crossovers between a time series and a threshold has similar settings with two time series crossover rules. The only difference is that one of the time series is comprised of a constant value representing a threshold. Typical technical indicators will fall into this category if it has a certain range of values. One example of the application is relative strength index (RSI) which is bounded between 0 and 100 which is interpreted as the momentum strength of the recent price change. When the RSI is over a threshold, a buy signal of 1 is created and a short signal of -1 will occur when the RSI is lower than the threshold.

$$signal_t = \begin{cases} 1 \text{ if } series1_t > threshold \\ 0 \text{ else} \end{cases} \quad (2)$$

*3) One time series and two threshold crossing*

Category 3 rules is a slight modification to the rules in category 2 where two thresholds were supplied to generate the trade signal. One of the thresholds is greater than the other. The typical setup of the trading rule is that when the time series of interest moves above the upper threshold, a buy or sell signal is generated, and a reverse signal would be generated when the time series moves below the lower threshold. When the time series is in between the higher and lower thresholds, a neutral signal, representing no money should be kept in the market, would be generated. The neutral position allows capital to be safely sitting on cash and stay away from excessive risks.

$$signal_t \begin{cases} 1 \text{ if } series1_t > threshold1 \\ -1 \text{ if } series1_t < threshold2 \\ 0 \text{ else} \end{cases} \quad (3)$$

*4) Three time series crossing*

Category 4 rules have three time series, like category 3 rules, with upper and lower thresholds replaced by upper and lower time series. One typical example is Bollinger bands where the upper band is always greater than the lower band at any time. A moving average, high or low prices could be used as the free moving time series. Category 4 rules also allows neutral position indication.

$$signal_t \begin{cases} 1 \text{ if } series1_t > series2_t \\ -1 \text{ if } series1_t < series2_t \\ 0 \text{ else} \end{cases} \quad (4)$$

*D. Sharpe and Stirling ratio (SSR)*

For optimisation of the trading rule parameters, the scoring algorithm and assumptions need to be stated. The metric used to compare the performance of each trading rule when applied to the training dataset. Without taking transaction costs into consideration, the currency pair is assumed to be purchased or sold at the close price of the latest candle. The overall return could be calculated as in (5) (6).

$$r_{total} = \sum signal_t \times log(close_t / close_{t-1}) \quad (5)$$

$$r_{total} = \sum_t s_t \times r_t \quad (6)$$

Where s represents signals and r represents log returns.

However, using the return as the performance measure is inadequate. To adjust for the risk factors, conventional methods use Sharpe ratio, which penalizes high standard deviation of the return, or Sterling ratio, which penalizes higher maximum of drawdown. Sharpe ratio penalizes negative returns less when the average return is generally low, and it would also penalize occasional high returns. Sterling ratio penalizes the maximum drawdown which could be experienced when the market changes suddenly and unexpectedly, which is very common in forex market. Hence, we propose a new ratio which incorporates the variance and drawdown factors in a single equation as in (7) and the name of the ratio is Sharpe and Stirling ratio (SSR) since it is derived from Sharpe ratio and Stirling ratio. In (7), the dot operator represents dot product and the cross operator represents multiplication between two scalars.

$$SS\ ratio = \frac{\sum_t s_t \times r_t}{\sigma(s_t \times r_t) \times |\sum_t s_t \times r_t, \; \forall \; s_t \times r_t < 0|} \quad (7)$$

The new ratio penalizes higher standard deviation as well as higher occurrence and values of drawdowns. The numerical value of the ratio could be interpreted as the proposed measure of risk-adjusted return. The reason for using the sum of drawdowns instead of maximum drawdown is to reduce the outlier effect that is unavoidable in a volatile market. Reducing the sum of drawdown experienced would also force the strategy to find a better strategy to minimize loss. The standard deviation of the returns would force the strategy to find a consistent trading strategy.

*E. Grid search for trading rule parameters*

Each trading rule would require parameters for associated technical indicators such as the moving average period size and thresholds for buy or sell limits. The parameters need to be tuned to achieve better performance on the trading system. A grid-search method is implemented and the search space for each individual rule contains at most 3 parameters which could be completed in polynomial time. An example is that the moving average window size must be integers and a range from 1 to 100 is selected for searching that indicates up to 100 previous price information could be incorporated into the trading rule. The scores are calculated by the SS ratio on the training dataset if the trading rule is implemented. The best scores and the corresponding parameters are found using the algorithm.

---

**Algorithm 1** Grid search parameter optimization

**START**

Setup trading rule functions

Initialize parameter list for each rule

**For** each rule **do**

    **For** each parameter in parameter list **do**

        Compute the SS ratio over data with parameter

        Save results to list

    **End for**

    Find the highest SS ratio and its parameter

**End for**

---

Fig. 4. Pseudocode for grid search algorithm

As a result of parameter optimization, 16 new features are generated that represents the trading signal based on each single trading rule. A sample of the engineered is shown in Fig. 5. A dataframe contains those 16 features will be optimized using machine learning algorithms to improve the performance of the overall trading system. The number of features is relatively low providing reasonable representation of the historical prices since the rules are based on technical indicators. The performance of the abstraction of the features will be tested using genetic algorithm.

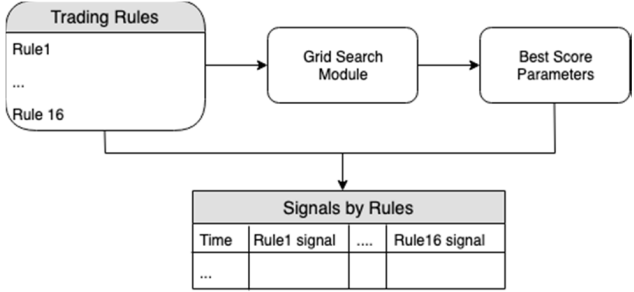

Fig. 5. One sample of the 16 engineering features

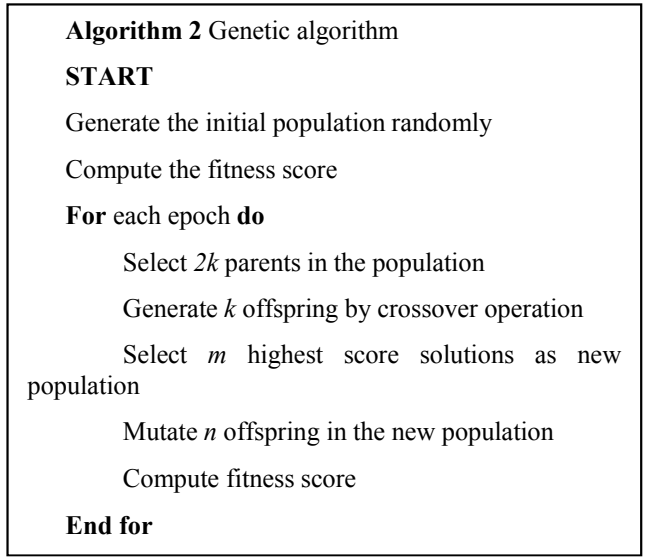

Fig. 6. Parameter optimization process

### F. Genetic algorithm module

Genetic algorithm mimics the natural selection process where the fittest individual will be selected and is usually applied to optimization problems. The steps in genetic algorithm involve selection, crossover and mutation stages. The pseudo-code for the genetic algorithm is included below.

---

**Algorithm 2** Genetic algorithm

**START**

Generate the initial population randomly

Compute the fitness score

**For** each epoch **do**

    Select *2k* parents in the population

    Generate *k* offspring by crossover operation

    Select *m* highest score solutions as new population

    Mutate *n* offspring in the new population

    Compute fitness score

**End for**

---

Fig. 7. Pseudocode for genetic algorithm

The problem setup is to use genetic algorithm to find a trading strategy that maximizes the performance. We will find a mapping of features S to the position of volume in the market *v*. The chromosomes for the genetic algorithm is a vector *w* that represents the weight for all features. Using linear mapping, the return over the trading period can be calculated as (9).

$$v_t = f(s_{1,t}, s_{2,t}, \dots, s_{n,t}) \quad (8)$$

$$v_{t+1} = \sum_i w_i \times s_{i,t} \quad (9)$$

The vector *v* is adjusted to have a maximum absolute value of 1 to represent the position as a percentage of the capital. A value of 0 means sitting on cash away from the market and values of 1 and -1 means long position and short position with 100% of the capital respectively. The output vector *v* represents a time series of positions in the market. The changing volume mechanism allows better risk control as the algorithm could achieve lower standard deviation by placing lower positions at volatile markets.

#### 1) Benchmark model: maximizing the return (GA-MR)

To achieve profit maximization, the conventional method is to compute the accumulated returns over the entire training period. Maximizing the overall return of the trading strategy will yield the solution. Equation (10) (11) is used as fitness function to maximize the overall return of the trading system. The maximizing return approach is used for comparison with the proposed fitness function.

Maximize

$$fit = \sum_t v_t \times r_t \quad (10)$$

Subject to

$$max \ |v_t| \leq 1 \quad (11)$$

#### 2) Proposed model: maximising the SSR (GA-MSSR)

Minimizing the risk without consideration for profit will not lead to good trading results. Hence, we propose using SSR as the proxy for the risk-adjusted return as shown in (12) (13). The use of SSR as the fitness score enables profit maximization with a certain level of risk management. The proposed method using the fitness function is named Genetic Algorithm Maximizing SS Ratio (GA-MSSR).

Maximize

$$fit = \frac{\sum_t v_t \times r_t}{\sigma(v_t \times r_t) \times |\sum_t v_t \times r_t, \ \forall \ v_t \times r_t < 0|} \quad (12)$$

Where $v_t$ is volume, $r_t$ is rate of return and $\sigma$ is standard deviation; subject to:

$$max \ |v_t| \leq 1 \quad (13)$$

## III. EXPERIMENT SETUP

### A. Data Collection

Historical data of 6 major currency pairs are tested including EURUSD, GBPUSD, AUDUSD, USDJPY, USDCAD and USDCHF. The timeframe and period used are 5 minutes data. from January 2018 to December 2019 which includes around 200,000 timesteps. The period of data is selected to only include the recent years that prevents potential regime changes in the market over time. 2018 data was used for training and 2019 data was used for testing.

### B. Cross-validation

The entire data is split with a ratio of 50% to obtain the training dataset and test dataset. To prevent data leaks, the first half of the entire period is used for training and the rest is retained for testing without random shuffling. This ensures that no future information would be used in the training data thus affecting the validity of the results on test data.

## C. Evaluation metrics

ROI: Return on investment is calculated as the annualized return of the trading system on the data. Higher returns are preferable.

$$ROI = \left(\frac{final\ balance}{initial\ balance}\right)^{\frac{365}{trading\ days}} - 1$$

SR: Sharpe ratio is calculated as the mean excess return over the standard deviation of the excess returns. Higher Sharpe ratio represents better return when the risk is adjusted. Therefore a higher Sharpe ratio represents a superior trading strategy.

$$SR = \frac{Accumulated\ return}{std(daily\ return)} \quad (11)$$

MD: Maximum drawdown is the lowest level of capital during the backtesting stage in percentage. A lower maximum drawdown also represents lower risks and uncertainties associated with the trading strategy.

$$MD = \min(balance) \quad (12)$$

AP: Average position is the average amount of capital invested in the market of the entire capital in percentage. A value of 1 represents 100% of capital is invested in the market all the time whereas 0.1 means 10% of the capital is invested in the market on average and the other 90% of capital is held as cash in the account without risks. Lower AP with high return could mean that the trading strategy implies that only a portion of capital is needed to generate a good return and the other portion of the capital is available to be invested on other assets.

$$AP = mean\left(\frac{volumes}{balance}\right) \quad (13)$$

## D. Parameter settings

The hyperparameters for the genetic algorithm are set as following:

Solution per population = 10

Number of parents mating = 4

Number of generation = 200

Mutation probability = 0.5

Crossover probability = 0.4

## E. Benchmark model

The simple buy and hold (B&H) and simple sell and hold (S&H) strategies are used as the benchmark system in the trading system. The genetic algorithm maximizing the return model (GA-MR) will also be compared with the proposed GA-MSSR model.

## F. Trading simulations

A 1:1 leverage is used for all models. We assume the forex security will only be bought or sold at the close price of the latest candle. We also assumed 0 transaction fees.

## IV. RESULTS AND DISCUSSION

### A. Results

The model performance evaluation results are shown in Table 1. It is observed in training dataset, naïve buy and hold (B&H) or sell and hold (S&H) strategies have the worst performance. Results indicate that randomly buy or sell a currency pair in forex market is equivalent to gambling which no one could make profits in the long term.

The other models GA-MR and GA-MSSR are all tested with positive results for all currency pairs. The performance differs for different pairs but on general, the model proved it works for different currency pairs. The proposed GA-MSSR model outperformed benchmarks in Sharpe ratio, maximum drawdown and average position metrics for all pairs. The return value is lower than GAMR which is as expected, but the different margin is very small. The return after adjusted for risk for GAMR would be smaller than GA-MSSR which has a higher Sharpe ratio and significantly lower drawdowns.

The pair that has the best performance is USDCAD where the return for GA-MSSR is higher than GA-MR. Empirical evidence suggests that the model generalizes better on the test data for this specific pair. However, the worst performing pair is USDJPY which obtained a much lower annualized return and relatively high drawdowns.

TABLE I. PERFORMANCE OF 6 MAJOR CURRENCY PAIRS ON 2019 5-MINUTE DATA WITH NO LEVERAGE

|  | EURUSD | | | | GBPUSD | | | | AUDUSD | | | |
|---|---|---|---|---|---|---|---|---|---|---|---|---|
|  | ROI | SR | MD | AP | ROI | SR | MD | AP | ROI | SR | MD | AP |
| B&H | -0.48% | -0.16 | -3.62% | 1 | 4.27% | 0.76 | -4.27% | 1 | 0.69% | 0.17 | -4.34% | 1 |
| S&H | 0.48% | 0.159 | -0.23% | 1 | -4.27% | -0.8 | -5.53% | 1 | -0.69% | -0.17 | -1.57% | 1 |
| GA-MR | 10.75% | 6.09 | -0.16% | 0.93 | 13.01% | 4.32 | -0.13% | 0.55 | 17.56% | 4.58 | -0.32% | 0.92 |
| **GA-MSSR** | 9.68% | **6.68** | **-0.03%** | **0.35** | 12.98% | **4.85** | **-0.08%** | **0.37** | 15.95% | **5.97** | **-0.13%** | **0.45** |
|  | USDCAD | | | | USDCHF | | | | USDJPY | | | |
|  | ROI | SR | MD | AP | ROI | SR | MD | AP | ROI | SR | MD | AP |
| B&H | -0.97% | -0.32 | -0.99% | 1 | -2.21% | -0.6 | -2.21% | 1 | 1.03% | 0.21 | -3.41% | 1 |
| S&H | 0.97% | 0.32 | -1.84% | 1 | 2.21% | 0.59 | -1.53% | 1 | -1.03% | -0.21 | -1.69% | 1 |
| GA-MR | 14.20% | 8.21 | -0.03% | 0.61 | 12.05% | 7.29 | -0.12% | 0.46 | 5.12% | 2.67 | -0.14% | 0.56 |
| **GA-MSSR** | 14.98% | **8.85** | **-0.01%** | **0.42** | 11.74% | **7.53** | **-0.09%** | **0.31** | 5.04% | **2.91** | **-0.12%** | **0.32** |

The average position for GA-MSSR has significantly lower values than other benchmarks. It shows that the GA-MSSR can achieve similar level of returns with much less capital invested in the market. With a lower position in the market, the underlying risks is also reduced.

The table results have a leverage of 1:1 where the common leverage used in forex could range from 1:20 to 1:300. A 1:20 leverage is applied to the GA-MSSR results and the accumulated returns over time is plotted in Fig. 8. All currency pairs are shown and the USDJPY has the worst performance. All other pairs showed a stable and consistent upward trend with minor drawdowns. The bad performance of USDJPY could be due to regime changes and undetected news events.

### B. Discussion

We propose the GA-MSSR model that can generate a superior trading strategy for 5-minute intraday data. The high ROI and Sharpe ratio indicates that the features derived using trading rules are good proxies to measure the trend of the price movement in the long term. The size of the feature space could grow infinitely large and the effect of feature space size is not covered in the paper. However, the mechanism of the proposed model differs from other researches. Most of the papers reviewed utilized directional prediction with machine learning or next day price prediction. The proposed model does not perform any prediction on the future market. Instead, it finds a set of rules and formulate strategies only according to the technical indicators rules to optimize the overall performance of the trading strategy. The directional prediction may only reveal the short-term trend and offer no information on the risk management side. With regression networks, accurately predicting the price at the next bar is very hard and longer timestep prediction has much higher variances. Trading rules are designed and optimized to ignore the short-term fluctuations and focus on the systematic practice to avoid risks and generate stable returns.

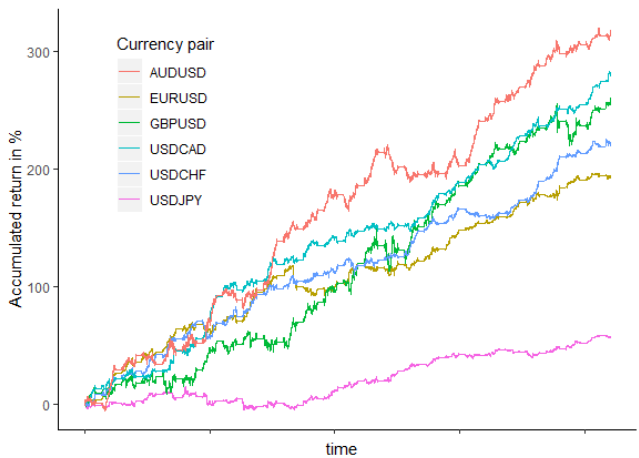

Fig. 8. Testing results accumulated return over time (2019) for all pairs tested with 1:20 leverage.

Sharpe ratio is one of the most important metrics to be considered when it comes to evaluating a trading strategy. Sharpe ratio measures the return over the standard deviation in the return. It is commonly known that with higher risks we could have higher returns. The Sharpe ratio stands for the gradient here. Therefore, the Sharpe ratio could also be interpreted as the "value for money" strength. Generally, the trading system with the higher the Sharpe ratio is a better choice. In a volatile market, such as forex, managing the risk is crucial to the long term management of a portfolio.

Leverage is a common method used in the forex market. It allows multiple amounts of capital to be invested whereas multiple loss or return would be experienced. The experiment implemented has no leverage which is 1:1 ratio and the Fig. 8 showed the simulated return using a 1:20 ratio. In practice, higher leverage means higher risks which is rewarded with multiplied returns. The limitation of using very high leverages is the chance of losing lots of money in one trade and the account would be blown. The GA-MSSR limited the maximum drawdown to low levels and it allowed moderate leverage to be applied.

## V. CONCLUSION, LIMITATIONS AND FUTURE WORK

The proposed feature engineering process provided quality features that are optimized based on the performance of the strategy. The work showed how the technical indicators could be combined and optimized for trading purposes. The proposed GA-MSSR RoboTrading system uses 16 features as a higher-level interpretation of the market. The proposed model used novel risk-adjusted return measure, SSR, as the fitness function and generated consistent and superior results compared to other benchmarks. The performance on the risk side is consistently good and the return achieved is outstanding. The best performing currency pair on the testing dataset is AUDUSD which obtained an annual return of 320% with a 1:20 leverage.

However, the proposed trading system has not taken the transaction fees into account due to different pricing policies by each broker and the time-variant nature of the spread cost. Therefore, accurately incorporating the transaction fee is a challenge to be conquered in the future. Other methods such as deep reinforcement learning [25] may be implemented to work with rule-based features to optimize portfolio performance for future works.